\documentclass[intlimits,twoside,a4paper]{article}

\usepackage{textcomp}
\usepackage{amsmath,amssymb}
\usepackage{graphicx}

\usepackage[T2A]{fontenc}
\usepackage[cp1251]{inputenc}

\usepackage[eqsecnum]{cmpj2}

\issue{2015}{18}{4}{43003}
\doinumber{10.5488/CMP.18.43003}

\title[Approximate solution for Fokker-Planck equation]%
{Approximate solution for Fokker-Planck equation
}
\author[M.T. Araujo, E. Drigo Filho]{M.T. Araujo\refaddr{label1,label2},
E. Drigo Filho\refaddr{label1,label3}}
\addresses{
\addr{label1} Instituto de Bioci\^{e}ncias, Letras e Ci\^{e}ncias Exatas, UNESP - Univesidade Estadual Paulista, \\
2265  Crist\'ov\~{a}o Colombo St., 15054-000, S\~{a}o Jos\'e do Rio Preto/SP, Brazil
\addr{label2} Institut f\"{u}r Physik, Universit\"{a}t Augsburg , Universit\"{a}tsstr. 1, D-86135, Augsburg, Germany
\addr{label3} Departamento de F\'{\i}sica Te\'orica, At\'omica y \'Optica, and IMUVA (Instituto de Matem\'aticas), \\ Universidad de Valladolid, 47011 Valladolid, Spain
}

\date{Received June 1, 2015, in final form October 6, 2015}
\authorcopyright{M.T. Araujo, E. Drigo Filho, 2015}

\begin{document}

\maketitle

\begin{abstract}
In this paper, an approximate solution to a specific class of the Fokker-Planck equation is proposed. The solution is based on the relationship between the Schr\"{o}dinger type equation with a partially confining and symmetrical potential. To estimate the accuracy of the solution, a function error obtained from the original Fokker-Planck equation is suggested. Two examples, a truncated harmonic potential and non-harmonic polynomial, are analyzed using the proposed method. For the truncated harmonic potential, the system behavior as a function of temperature is also discussed.
\pacs 05.10.Gg, 02.30.Mv
\end{abstract}


\section{Introduction}

Differential equations are used to model different natural phenomena such as diffusion that are of great importance in many physical, chemical and biological processes \cite{ref1}.

In general, diffusive processes can be treated via the Fokker-Planck equation \cite{ref2,ref3,ref4,ref5,ref6}. This equation is obtained from the Langevin equation and gives the probability of finding a given particle in a state ${x}$ at time ${t}$. The usual representation of the Fokker-Planck equation is given as follows:
\begin{equation}
\label{eq1}
\frac{{\partial P(x,t)}}{{\partial t}} =  - \frac{\partial }{{\partial x}}\left[ {f(x)P(x,t)} \right] + Q\frac{{{\partial ^2}P(x,t)}}{{\partial {x^2}}},
\end{equation}
where ${t}$ is the time variable, and ${x}$ is the variable characteristic of the system (which may be identified, for example, as velocity). ${Q}$ is the diffusion coefficient and ${P}({x,t})$ is the probability distribution. The ${f}({x})$ function is known as the external force that acts on the system, although this designation is only suitable when ${x}$ represents velocity. This function can be identified as the derivative of a potential function ${V}({x})$: ${f}({x}) = -\partial {V}({x})/\partial {x}$.

Different methods for treating the Fokker-Planck equation have been suggested as, for example, its association with a Schr\"{o}dinger type equation \cite{ref2,ref7}. However, there are only a few cases where the equation (\ref{eq1}) has an exact analytical solution.

In a recent study \cite{ref8}, an approximate solution for partially confining potentials was proposed, where part of the solution is written in terms of functions that arise from the solution of the Schr\"{o}dinger type equation for bound states and part is composed of functions originating from the free particle. This solution method is discussed in reference \cite{ref9} for the particular case of the Rosen-Morse potential. This potential has an exact solution to the Schr\"{o}dinger equation. As expected, it can be concluded that the results obtained using the functions derived from the original Schr\"{o}dinger type equation provide better solutions than those obtained by the proposed approximate method. However, this approach proves to be very restrictive, since the number of potentials with exact solutions to the Schr\"{o}dinger equation is very small.

In this paper, an improvement to the previously proposed solution \cite{ref8} is suggested, producing a more accurate result for cases of symmetrical partially-confining potentials. To check the validity of the results, the error on the results is estimated through direct substitution of the approximate solution into the Fokker-Planck equation.

In the study of the truncated harmonic potential, the amount of particles that escape from the potential well are calculated and a phase transition for the systems being studied is identified.

In section~\ref{sec2}, there is a brief discussion of the Fokker-Planck equation and the approximate solution method is presented, with the changes suggested in relation to the previously proposed solution \cite{ref8}.  In sections~\ref{sec3} and \ref{sec4}, the proposed method is applied to two partially-confining potentials. In the first case, the region of the well potential is described by a harmonic potential and, for the second case, the well is given by a non-harmonic polynomial potential. Finally, the conclusions are presented in section~\ref{sec5}.


\section{The Fokker-Planck equation}\label{sec2}

Given the importance and difficulty of solving the Fokker-Planck equation (FPE), different methods have been proposed to study this equation. Such methods include numerical treatments \cite{ref10,ref11} and mapping the FPE onto a  Schr\"{o}dinger type equation \cite{ref2,ref7,ref12}. In the latter case, the expression of ${P}({x,t})$ is given by a series of functions as follows:
\begin{equation}
\label{eq2}
P(x,t) = \sum\limits_{n = 0}^\infty  {{a_n}{\phi _0}(x){\phi _n}(x)} {\re^{ - t\left| {{\lambda _n}} \right|}},
\end{equation}
where the eigenfunctions $\phi_n({x})$ and the eigenvalues $\lambda_n$  are the solution of the Schr\"{o}dinger type equation obtained from the FPE (\ref{eq1}):
\begin{equation}
\label{eq3}
{\lambda_n}{\phi _n} =  - \frac{{{\phi _n}}}{2}\left\{ {\frac{{\rd f(x)}}{{\rd x}} + \frac{{f^2{{(x)}}}}{{2Q}}} \right\} + Q\frac{{{\rd^2}{\phi _n}}}{{\rd{x^2}}}.
\end{equation}

Comparing equation (\ref{eq3}) with the Schr\"{o}dinger equation, it can be seen that the term in parentheses is equivalent to a potential function. Thus, this term is commonly called the effective potential, $V_\textrm{ef}(x)$:
\begin{equation}
\label{eq4}
{V_\textrm{ef}}(x) = \frac{1}{2}\left\{ {\frac{{\rd f(x)}}{{\rd x}} + \frac{{f^2{{(x)}}}}{{2Q}}} \right\}.
\end{equation}

The probability distribution indicated by equation (\ref{eq2}) assumes that the initial condition for the time ${t} = 0$ is the probability distribution expressed by a delta function ${P}({x},0) = \delta({x})$.

In order to obtain the numerical solution shown in equation (\ref{eq2}), it is necessary to define a criterion to the cut off in the sum to truncate the series. Another problem arises when one cannot get the solution of the corresponding Schr\"{o}dinger equation (\ref{eq3}). To get around this last problem, an approximate analytical solution composed of two parts was suggested in a previous paper \cite{ref8}. This solution involves the discrete levels of the problem and the other one uses a Gaussian distribution for all continuous states. Although the solution proposed in \cite{ref8} agrees with the numerical results, the adopted approach can be improved upon.

The attention here is focused on specific issues involving symmetrical and partially confined potentials, where it is not possible to get the complete solution of the Schr\"{o}dinger type equation (\ref{eq3}). Thus, for these types of potential, a similar treatment to that of reference \cite{ref9}, for example, is unfeasible. For the potentials studied, there are regions where the spectrum is continuous and a region where there may be bound states (i.e., a discrete spectrum).  For ${x} > {d}$ and ${x} < - {d}$, it is assumed that the potential is constant and the spectrum is continuous. In $-{d} < {x} < d$, there is a potential well and the spectrum becomes discrete. It is also assumed that the potential is continuous, especially at the points ${x} = \pm {d}$ that correspond to the intersections between the region where the potential is constant and the well potential region (figures~\ref{fig1},  \ref{fig6} and  \ref{fig8} exemplify the type of the studied potential). The probability distribution in such cases is calculated separately for each region of the potential.
\begin{equation}
\label{eq5}
P(x,t) = \left\{ \begin{array}{ll}
{N_\textrm{I}}{\rho _\textrm{I}}(x,t), & \hbox{$ - d > x$},\\
{N_\textrm{II}}{\rho _\textrm{II}}(x,t), & \hbox{$ - d \leqslant x \leqslant d $}, \\
{N_\textrm{III}}{\rho _\textrm{III}}(x,t), & \hbox{ $x > d$}.
\end{array} \right.
\end{equation}

The function $\rho({x,t})$ in (\ref{eq5}) is the probability distribution for each region; ${N_\textrm{I}}$, ${N_\textrm{II}}$ and ${N_\textrm{III}}$ are related to the normalization in each piece of the probability distribution. For regions ({I} and {III}) of the continuous spectrum, the $\rho$ functions are given by a Gaussian function, as suggested previously \cite{ref8}
\begin{equation}
\label{eq6}
{\rho _\textrm{I}}(x,t) = {\rho _\textrm{III}}(x,t) = \frac{1}{{\sqrt {4Q\pi t} }}{\re^{ - {{{x^2}} \mathord{\left/
 {\vphantom {{{x^2}} {4Qt}}} \right.
 \kern-\nulldelimiterspace} {4Qt}}}}
\end{equation}
and for region {II}, the function $\rho$ is expressed by the series indicated in (\ref{eq2}) with a limited number of eigenvalues (${j}$)
\begin{equation}
\label{eq7}
{\rho _\textrm{II}}(x,t) = \sum\limits_{i = 0}^j {{a_i}{\phi _0}(x){\phi _i}(x)} {\re^{ - t\left| {{\lambda _i}} \right|}}.
\end{equation}

The number ${j}$ corresponds to the number of discrete eigenvalues present in the potential well under analysis.

As the functions $\rho_\textrm{I}({x,t})$ and $\rho_\textrm{III}({x,t})$ are equal, because of the symmetry of the problem, one can assume that the normalization parameters ${N_\textrm{I}}$ and ${N_\textrm{III}}$ are the same. It can also be assumed that at the interface points between the potentials ($\pm {d}$), the distribution $\rho({x,t})$, equation (\ref{eq5}) should be continuous, i.e., $\rho_\textrm{II}({d,t}) = \rho_\textrm{III}({d,t})$ and $\rho_\textrm{I}({-d,t}) = \rho_\textrm{II}({-d,t})$. Thus, the condition of the continuity of the distribution and the symmetry of the problem imply that ${N_\textrm{I}} = {N_\textrm{III}}$ and
\begin{equation}
\label{eq8}
{N_\textrm{II}}(t) = {N_\textrm{I}}\frac{{{\re^{ - {{{d^2}} \mathord{\left/
 {\vphantom {{{d^2}} {4Qt}}} \right.
 \kern-\nulldelimiterspace} {4Qt}}}}}}{{\sqrt {4\pi Qt} }}{\left\{ {\sum\limits_{i = 0}^j {{a_i}{\phi _0}(d){\phi _i}(d)} {\re^{ - t\left| {{\lambda _i}} \right|}}} \right\}^{ - 1}}.
\end{equation}

This relationship (\ref{eq8}) shows that the normalization ${N_\textrm{II}}$ depends on ${N_\textrm{I}}$ and also, in general, depends on time. To simplify the notation, we will rewrite equation (\ref{eq8}) as ${N_\textrm{II}}({t}) = {N_\textrm{I}} {g}({t})$, such that:
\begin{equation}
\label{eq9}
g(t) = \frac{{{\re^{ - {{{d^2}} \mathord{\left/
 {\vphantom {{{d^2}} {4Qt}}} \right.
 \kern-\nulldelimiterspace} {4Qt}}}}}}{{\sqrt {4\pi Qt} }}{\left\{ {\sum\limits_{i = 0}^j {{a_i}{\phi _0}(d){\phi _i}(d)} {\re^{ - t\left| {{\lambda _i}} \right|}}} \right\}^{ - 1}}.
\end{equation}

Therefore, the overall probability distribution for a problem with the discussed features (a partially confining and symmetric potential) is obtained by equation (\ref{eq5}), subject to the condition (\ref{eq8}), i.e.,
\begin{equation}
\label{eq10}
P(x,t) = \left\{ \begin{array}{l}
 {N_\textrm{I}}\frac{1}{{\sqrt {4Q\pi t} }}{\re^{ - {{{x^2}} \mathord{\left/
 {\vphantom {{{x^2}} {4Qt}}} \right.
 \kern-\nulldelimiterspace} {4Qt}}}}, \\
 {N_\textrm{I}}g(t)\sum\limits_{i = 0}^j {{a_i}{\phi _0}(x){\phi _i}(x)} {\re^{ - t\left| {{\lambda _i}} \right|}}, \\
 {N_\textrm{I}}\frac{1}{{\sqrt {4Q\pi t} }}{\re^{ - {{{x^2}} \mathord{\left/
 {\vphantom {{{x^2}} {4Qt}}} \right.
 \kern-\nulldelimiterspace} {4Qt}}}}, \\
 \end{array} \right.\begin{array}{*{20}{l}}
   \begin{array}{l}
  - d > x, \\
  \\
 \end{array}  \\
   { - d \leqslant x \leqslant d},  \\
   \begin{array}{l}
  \\
 x > d. \\
 \end{array}
\end{array}
\end{equation}
Applying the normalization condition to the probability distribution (\ref{eq10}), one gets
\begin{equation}
\label{eq11}
{N_\textrm{I}}(t) = {{\left\{ {\frac{2}{{\sqrt {4\pi Qt} }}\int\limits_d^\infty  {{\re^{ - {{{x^2}} \mathord{\left/
 {\vphantom {{{x^2}} {4Qt}}} \right.
 \kern-\nulldelimiterspace} {4Qt}}}}} \rd x + g(t)\int\limits_{ - d}^d {\sum\limits_{i = 0}^j {{a_i}{\phi _0}(x){\phi _i}(x)} {\re^{ - t\left| {{\lambda _i}} \right|}}} \rd x } \right\}^{-1}}}.
\end{equation}

According to this result, the normalization parameter $N_\textrm{I}$ is dependent on time and the probability distribution (\ref{eq10}) should be normalized for each time value. Looking at the approximate solution given by equation (\ref{eq10}), one can see that when ${d}$ is close to zero the system approximated to a free particle system and the probability distribution approximates to a Gaussian. On the other hand, when ${d}$ is very large (${d} \to \infty$), i.e., the size of the system tends to an infinite well, the solution of the problem is given by the usual discrete series of functions, as shown in equation (\ref{eq2}).

The difference between the solution given by (\ref{eq10}) and that shown in reference \cite{ref8} is that here, the Gaussian is used only in areas where the potential is constant, while in reference \cite{ref8}, it was suggested that it should be included in all regions of the space. Numerical results show that the new approach is more suitable, leading to more accurate results.

In general, when there is a partial confinement, a temporal dependency arises already in the coefficient ${N_\textrm{I}}$ . Under these conditions, the probability distribution for large times in the region of the potential well (region {II}) may go to zero, which indicates the escape of particles from the minimum region of the potential.

From the proposed solution it is suggested that the escape of particles from the potential well can be quantified by the value ${Y}({t,Q})$ defined by:
\begin{equation}
\label{eq12}
Y(t,Q) = {N_\textrm{I}}(t)g(t)\int\limits_{ - d}^d {\sum\limits_{i = 0}^j {{a_i}{\phi _0}(x){\phi _i}(x)} {\re^{ - t\left| {{\lambda _i}} \right|}}} \rd x.
\end{equation}

The function ${Y}({t,Q})$ gives the number of particles within the confinement region for each time ${t}$ and different values of the diffusion coefficient (${Q}$). The functions ${N_\textrm{I}}({t})$ and ${g}({t})$ are given by expressions (\ref{eq11}) and (\ref{eq9}), respectively.

Equation (\ref{eq12}) also allows the evaluation of the influence of temperature in the escape process of the particles in the well. Assuming that the temperature is proportional to the diffusion coefficient \cite{ref2}, the calculation of the population for different values of ${Q}$ allows for the analysis of the evolution of the system in terms of temperature. This information allows, in principle, the study of the thermodynamic properties of the system, such as phase transitions.

To check the accuracy of the proposed method, we introduce the function $\varepsilon({x,t})$  based on the Fokker-Planck equation (\ref{eq1})
\begin{equation}
\label{eq13}
\varepsilon (x,t) = \frac{{\partial P(x,t)}}{{\partial t}} - \left\{ { - \frac{\partial }{{\partial x}}\left[ {f(x)P(x,t)} \right] + Q\frac{{{\partial ^2}P(x,t)}}{{\partial {x^2}}}} \right\}.
\end{equation}

This function provides a quantitative parameter to check if the solution approximates to the actual solution of the problem. If the solution is accurate, then $\varepsilon ({x,t}) = 0$. In this way, the further this function $\varepsilon({x,t})$ is close to zero, the better is the proposed function ${P}({x,t})$ to describe the real solution for the system under study.

Substituting the solution presented in (\ref{eq10}) into expression (\ref{eq13}), it can be seen that for ${x} < - {d}$ and ${x} > {d}$, where ${f}({x})$ is zero, we obtain
\begin{equation}
\label{eq14}
{\varepsilon _\textrm{I}}(x,t) = \frac{1}{{\sqrt {4Q\pi t} }}{\re^{ - {{{x^2}} \mathord{\left/
 {\vphantom {{{x^2}} {4Qt}}} \right.
 \kern-\nulldelimiterspace} {4Qt}}}}\frac{{\rd{N_\textrm{I}}(t)}}{{\rd t}},
\end{equation}
and for $- {d} < {x} < {d}$, the expression $\varepsilon ({x,t})$ is obtained by direct substitution of the solution in this region [equation (\ref{eq10})] in equation (\ref{eq13}).

Since the construction of $\varepsilon ({x,t})$ involves derivatives of the probability distribution ${P}({x,t})$, it is worth noting that, close to the points where the derivative is discontinuous,  the use of this criterion is impaired and should be used with caution. Thus, in this study, $\varepsilon ({x,t})$ was not defined for values of ${x}$ close to
$\pm {d}$. However, the use of the function $\varepsilon ({x,t})$ to evaluate the solution avoids comparisons with solutions obtained by other methods which could, in itself, introduce an additional error.


\section{Truncated harmonic oscillator}
\label{sec3}

In this section, we apply the approximate solution of the FPE to a model with a truncated harmonic oscillator, whose strength is given by
\begin{equation}
\label{eq15}
f(x) = \left\{ \begin{array}{l}
  - kx, \\
 0,  \\
 \end{array} \right.\begin{array}{*{20}{l}}
   { - d \leqslant x \leqslant d},  \\
   { - d > x{\quad \text{and} \quad } x > d}  \\
\end{array}
\end{equation}
with ($\pm {d}$) being the interface points between the harmonic potential and the constant potential. Substituting this expression of force into equation (\ref{eq4}), it can be seen that the effective potential can be identified by
\begin{equation}
\label{eq16}
V(x) = \left\{ \begin{array}{l}
 {v_0}, \\
 \frac{{{k^2}{x^2}}}{{4Q}} - \frac{k}{2}, \\
 \end{array} \right.{\rm{         }}\begin{array}{*{20}{l}}
   { - d > x{\quad \text{and}\quad}x > d},  \\
   { - d \leqslant x \leqslant d},  \\
\end{array}
\end{equation}
where ${v_0} = k^2 d^2/4Q - k/2$ is a constant value chosen such that the potential function is continuous.

\begin{figure}[!t]
\centerline{
\includegraphics[width=0.5\textwidth]{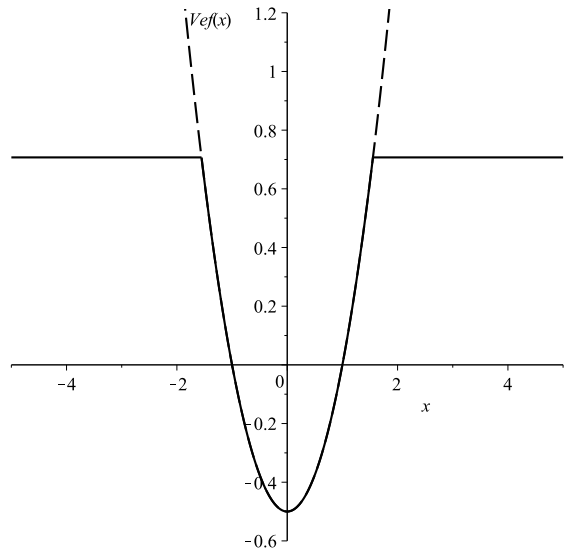}
}
\caption{Comparative graph of the truncated harmonic potential (solid line) and the usual harmonic potential (dashed line) for ${Q} = 1$, ${k} = 1.4$,  ${d} = 1.5$.} \label{fig1}
\end{figure}

Figure~\ref{fig1} shows the harmonic potential graph (dotted line) and the truncated harmonic potential, equation (\ref{eq16}) (continuous line). For the potential studied, the solid line in figure~\ref{fig1} shows that there are two distinct regions: one is a potential well region described by a harmonic potential and the other is described by a constant potential.

The harmonic oscillator problem is a case in which equation (\ref{eq3}) has a full analytical solution \cite{ref2} and the probability distribution is given by:
\begin{equation}
\label{eq17}
P(x,t) = \sum\limits_{n = 0}^\infty  {\left( {\frac{1}{{{2^n}n!}}\sqrt {\frac{\alpha }{\pi }} } \right){\re^{ - \alpha {x^2}}}{H_n}\left( {\sqrt \alpha  x} \right){H_n}(0){\re^{ - t\left| {{\lambda _n}} \right|}}} ,
\end{equation}
where ${H_n}$ are Hermite polynomials, $\alpha = {k/2Q}$ and ${k}$ is a constant. The eigenvalues ${\lambda_n}$ are equal to ${nk}$ (${n} = 0, 1, 2, \ldots, \infty$). Therefore, replacing the function ${f}({x})$, equation (\ref{eq15}), in the FPE (\ref{eq3}), using the approach presented in the previous section, the proposed solution for this example is given by:
\begin{equation}
\label{eq18}
P(x,t) \!=\! \left\{ \begin{array}{ll}
 \!{N_\textrm{I}}\frac{1}{{\sqrt {4Q\pi t} }}{\re^{ - {{{x^2}} \mathord{\left/
 {\vphantom {{{x^2}} {4Qt}}} \right.
 \kern-\nulldelimiterspace} {4Qt}}}}, & \hbox{$- d > x{\quad \text{and} \quad}x > d$},\\
 \!{N_\textrm{I}}g(t)\!\sum\limits_{n = 0}^j {\left( {\frac{1}{{{2^n}n!}}\sqrt {\frac{\alpha }{\pi }} } \right){\re^{ - \alpha {x^2}}}\!{H_n}\left( {\sqrt \alpha  x} \right)\!{H_n}(0){\re^{ - t\left| {{\lambda _n}} \right|}}}, & \hbox{${ - d \leqslant x \leqslant d} $} \\
 \end{array} \right.
\end{equation}
with ${j}$ being the maximum number of discrete states in the potential well region. The function ${g}({t})$ in this case is found by substituting the discrete functions of equation (\ref{eq18}) in equation (\ref{eq9}) and thus
\begin{equation}
\label{eq19}
g(t) = \frac{{{\re^{ - {{{d^2}} \mathord{\left/
 {\vphantom {{{d^2}} {4Qt}}} \right.
 \kern-\nulldelimiterspace} {4Qt}}}}}}{{\sqrt {4\pi Qt} }}{\left\{ {\sum\limits_{n = 0}^j {\left( {\frac{1}{{{2^n}n!}}\sqrt {\frac{\alpha }{\pi }} } \right){\re^{ - \alpha {x^2}}}{H_n}\left( {\sqrt \alpha  x} \right){H_n}(0){\re^{ - t\left| {{\lambda _n}} \right|}}} } \right\}^{ - 1}}
\end{equation}
and ${N_\textrm{I}}({t})$ is obtained by normalization, equation (\ref{eq11}),
\begin{equation}
\label{eq20}
{N_\textrm{I}}(t) = {{\left\{ {\frac{2}{{\sqrt {4\pi Qt} }}\int\limits_d^\infty  {{\re^{ - {{{x^2}} \mathord{\left/
 {\vphantom {{{x^2}} {4Qt}}} \right.
 \kern-\nulldelimiterspace} {4Qt}}}}} \rd x + g(t)\int\limits_{ - d}^d {\sum\limits_{i = 0}^j {\left( {\frac{1}{{{2^i}i!}}\sqrt {\frac{\alpha }{\pi }} } \right){\re^{ - \alpha {x^2}}}{H_i}\left( {\sqrt \alpha  x} \right){H_i}(0){\re^{ - t\left| {{\lambda _i}} \right|}}} } \rd x} \right\}^{-1}}}.
\end{equation}

It is assumed that, within the well, the truncation of the potential only slightly alters the original solutions of the harmonic oscillator. Thus, for the region between
$- {d} \leqslant  {x} \leqslant {d}$, the eigenfunctions [equation (\ref{eq17})] and eigenvalues ($\lambda_n = {nk}$) are the same as for the harmonic potential. The only difference is that the number of terms of the series was limited taking into account the height of the potential well.

\begin{figure}[!t]
\centering{
\includegraphics[height=7cm]{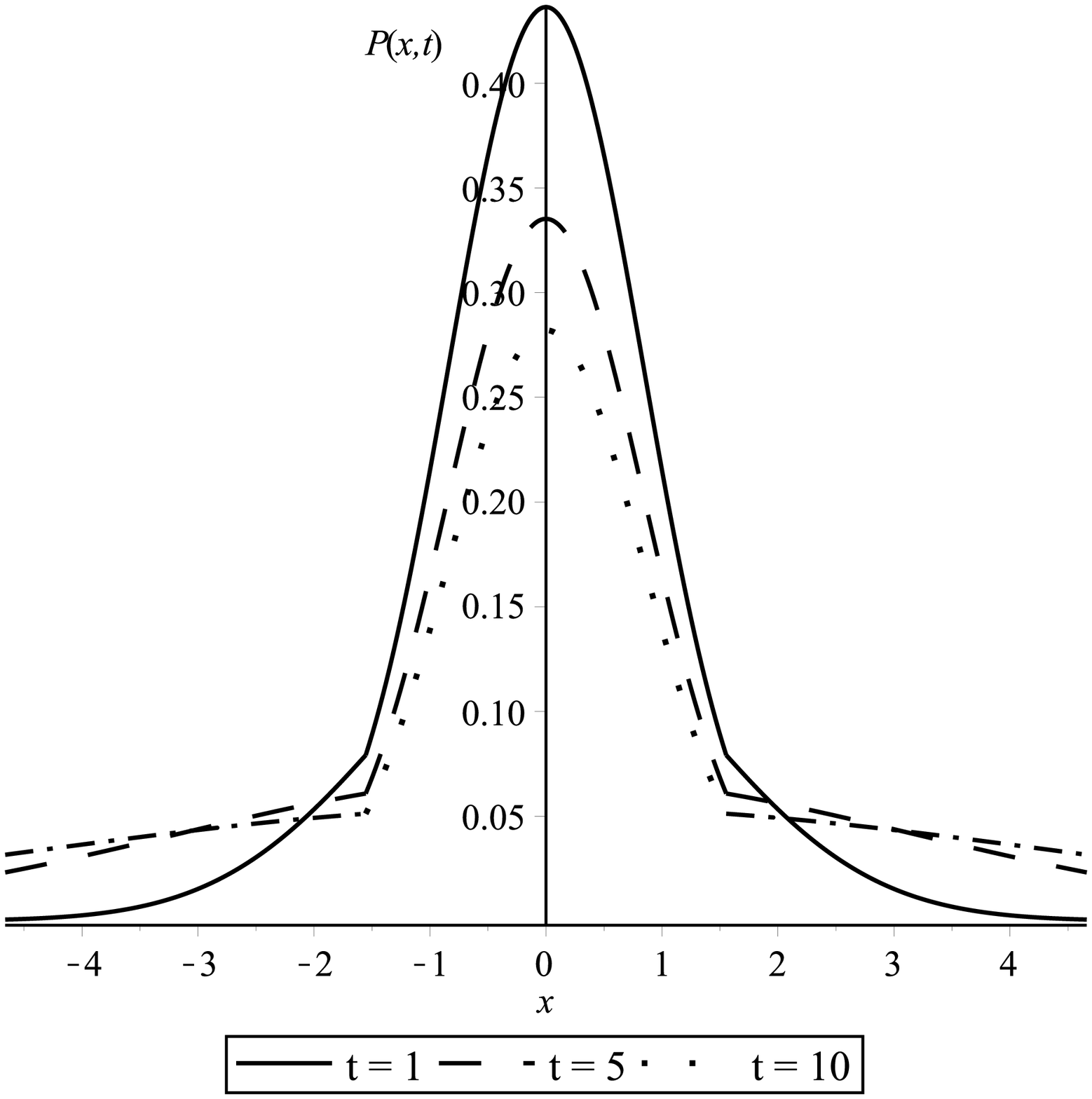}
}
\quad 
{
\includegraphics[height=7cm]{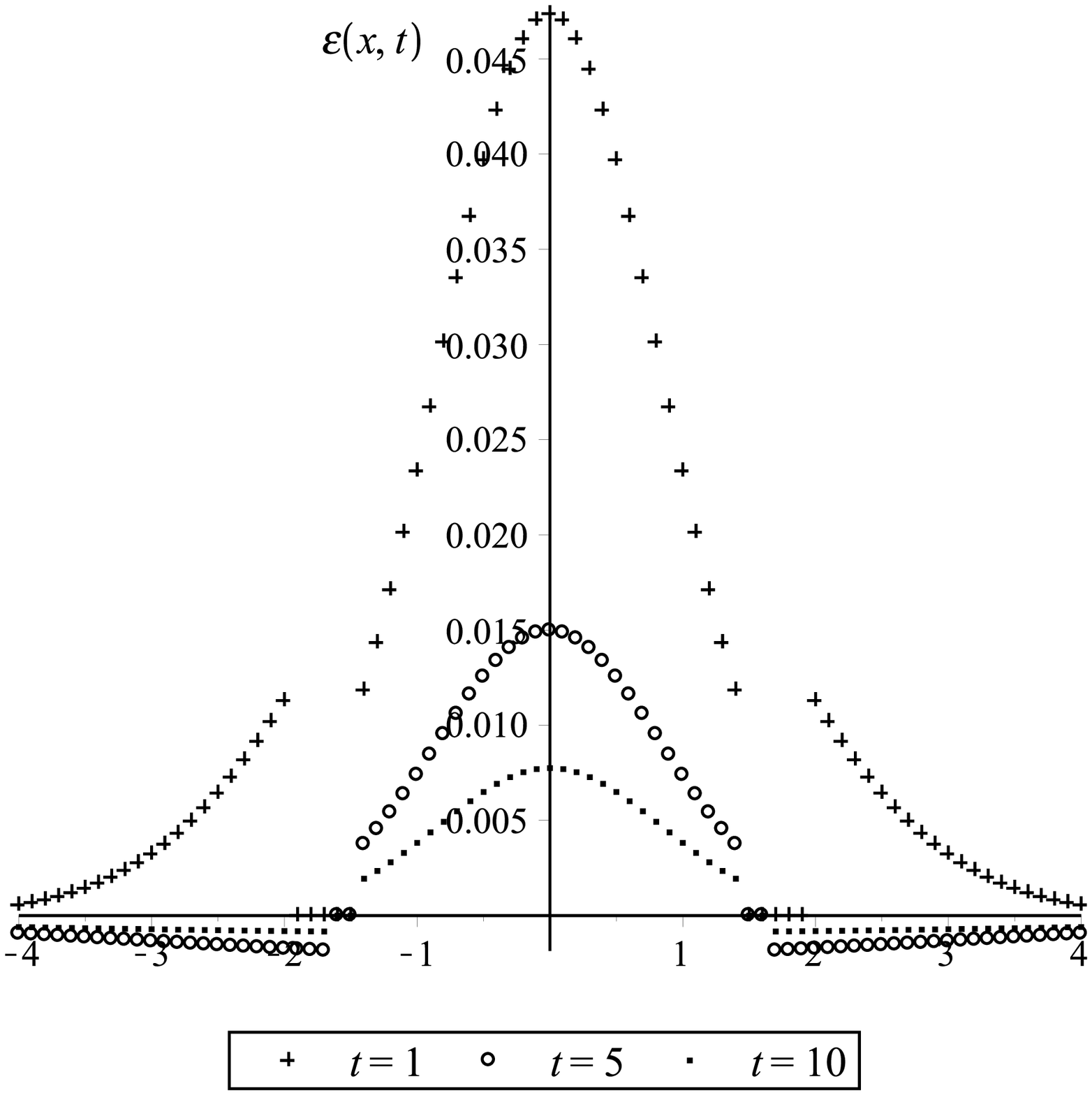}
}
\centerline{(a) \hspace{0.47\textwidth} (b)}
\caption{(a) The probability distribution (\ref{eq18}) versus ${x}$ for different values of time. The parameters used are: ${v_0} = 0.5$, ${d} = 1.55$, ${k} = 1.4$,  ${Q} = 1$. (b) Estimated error $\varepsilon ({x,t})$ for each solution.}
\label{fig2}
\end{figure}

Figure~\ref{fig2} shows the approximate probability distribution (\ref{eq18}) for the truncated harmonic potential [figure~\ref{fig2}~(a)] and the error associated with this solution given by the function $\varepsilon ({x,t})$ [figure~\ref{fig2}~(b)], equation (\ref{eq13}), for different values of time. In the construction of figure~\ref{fig2}, the values for the constants ${k} = 1$, ${Q} = 1$, ${d} = 1.55$  ${v_0} = 0.5$ were used. In this example there is only one discrete level in the well with an eigenvalue of zero, $\lambda_0 = 0$.

Observing the figure~\ref{fig2}~(a), one can see that the probability distribution is greater in the region of minimum potential, even for extended periods of time. Since the given solution was constructed using an approximate method, one can see from figure~\ref{fig2}~(b) that the probability distribution (\ref{eq18}) does not completely satisfy the Fokker-Planck equation, in other words, $\varepsilon ({x,t}) \ne  0$. However, it is noted that the calculated errors are small and decrease as time increases. The larger relative errors appear when the probability distribution is calculated within the region of the well potential.

In the vicinity of the interface points ($\pm d$), the error of the solution shows a discontinuity. This discontinuity is expected since the potential behavior studied is composed by joining different functions, and their profile (figure~\ref{fig1}) is not smooth for the whole curve.

\begin{figure}[!t]
\centerline{
\includegraphics[width=0.48\textwidth]{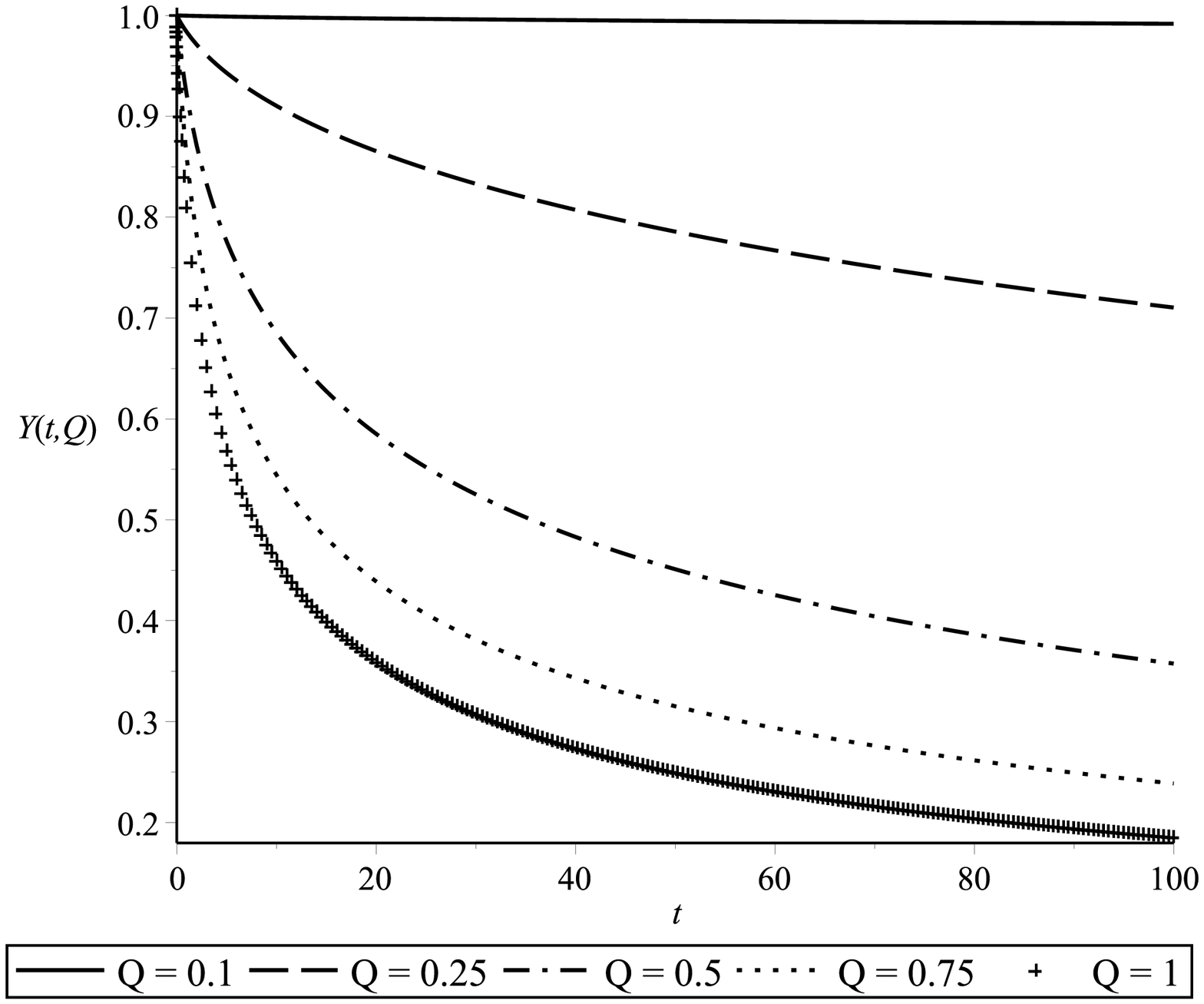}
\hspace{2mm}
\includegraphics[width=0.49\textwidth]{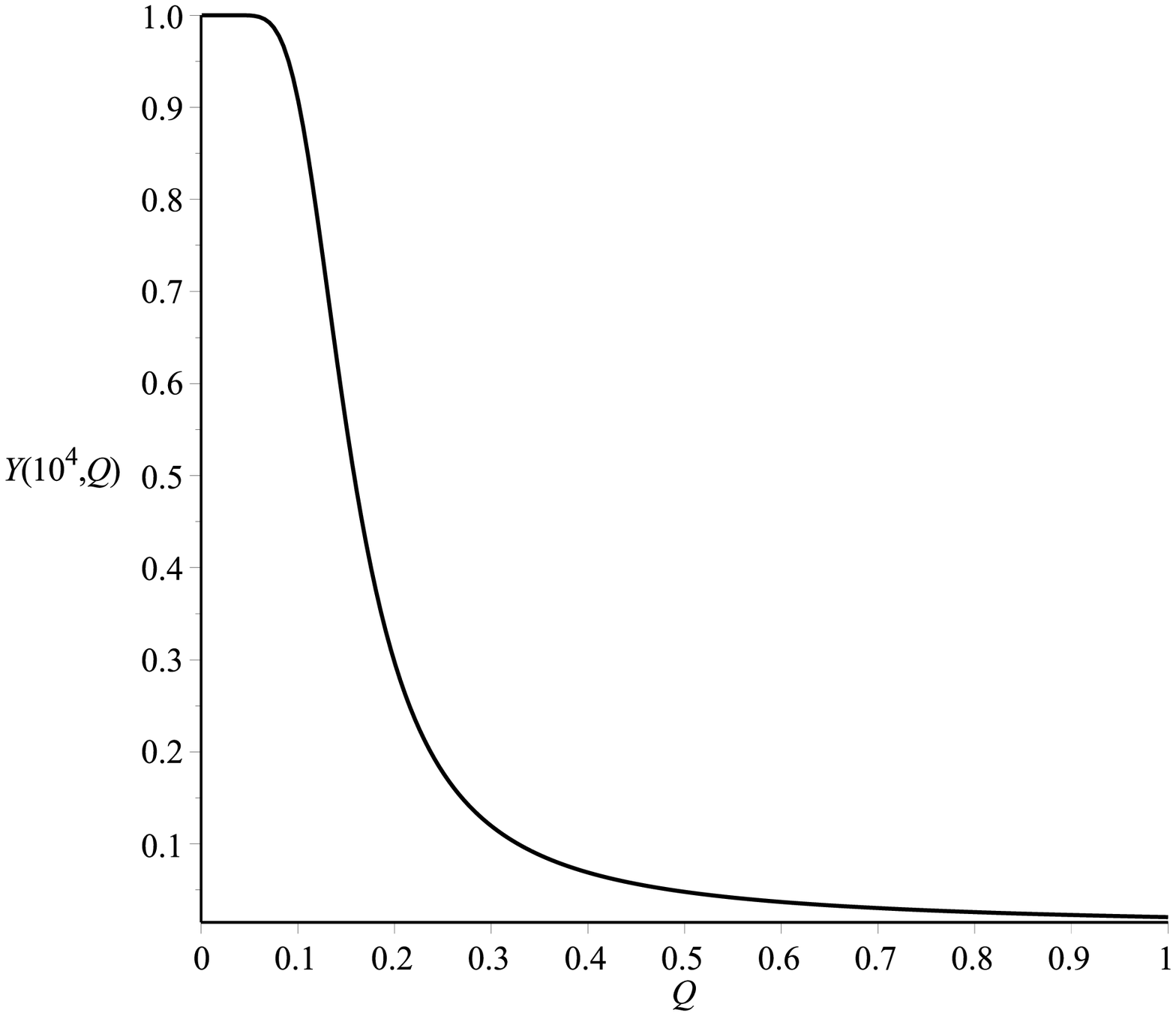}
}
\parbox[t]{0.49\textwidth}{
\caption{Curves describing the population near the minimum of potential, equation (\ref{eq12}) versus time ${t}$ for ${k} = 1.4$ and the well depth ${v_0} = 0.5$.} \label{fig3}
}
\parbox[t]{0.49\textwidth}{
\caption{Curve of ${Y}({t,Q})$, equation (\ref{eq12}) versus ${Q}$ for the truncated harmonic oscillator with only one state within the well [equation (\ref{eq10}) with ${j} = 0$] and for a large value of time (${t} = 10^4$).} \label{fig4}
}
\end{figure}

Through the probability distribution ${P}({x,t})$, equation (\ref{eq18}), the variation of the number of particles in the region of minimum potential can be calculated, equation (\ref{eq12}). The curves in figure~\ref{fig3} show the variation in the number of particles in the region of the potential well for different values of the diffusion coefficient.

In the definition of the potential used, equation (\ref{eq16}), the depth of the well (related to ${v_0}$) depends on the diffusion coefficient (${Q}$) and the interface point (${d}$).Thus, to maintain the fixed value of ${v_0}$ (equal to 0.5) for different values of ${Q}$, it is necessary to change the value of ${d}$. Maintaining a fixed value ${v_0}$ ensures that, within the potential well, the number of eigenvalues ${j}$ that are solutions of the Schr\"{o}dinger type equation (\ref{eq3}) is fixed. In the example discussed here, ${v_0} = 0.5$, there is only one eigenvalue of this kind.

Since it is assumed that the diffusion coefficient is proportional to temperature \cite{ref2}, lower values of ${Q}$ represent lower system temperatures.  Figure~\ref{fig3} shows that the decrease in the population of the region of the potential well depends on the value of ${Q}$, i.e., it depends on the temperature.

Initially the number of particles in the region of the well has the maximum value and with the increase in time this number of particles decreases. This drop in the number of particles is more pronounced for larger values of ${Q}$. This behavior is expected and consistent with the behavior of a system subject to a non-confining potential.

Figure~\ref{fig4} represents the behavior of the function ${Y}({t,Q})$, equation (\ref{eq12}), for a very large time value (${t} = 10^4$). In this figure, it can be seen that for small values of ${Q}$ (typically ${Q}$ lower than 0.1), the population is confined to the minimum potential region. On the other hand, for larger values of ${Q}$, the number of particles within the well of potential decreases to zero. This behavior shows a phase transition in which the particles remain in the well of potential at low temperatures, and at high temperatures the potential well becomes emptied.

\begin{figure}[!b]
\centering{
\includegraphics[height=7.25cm]{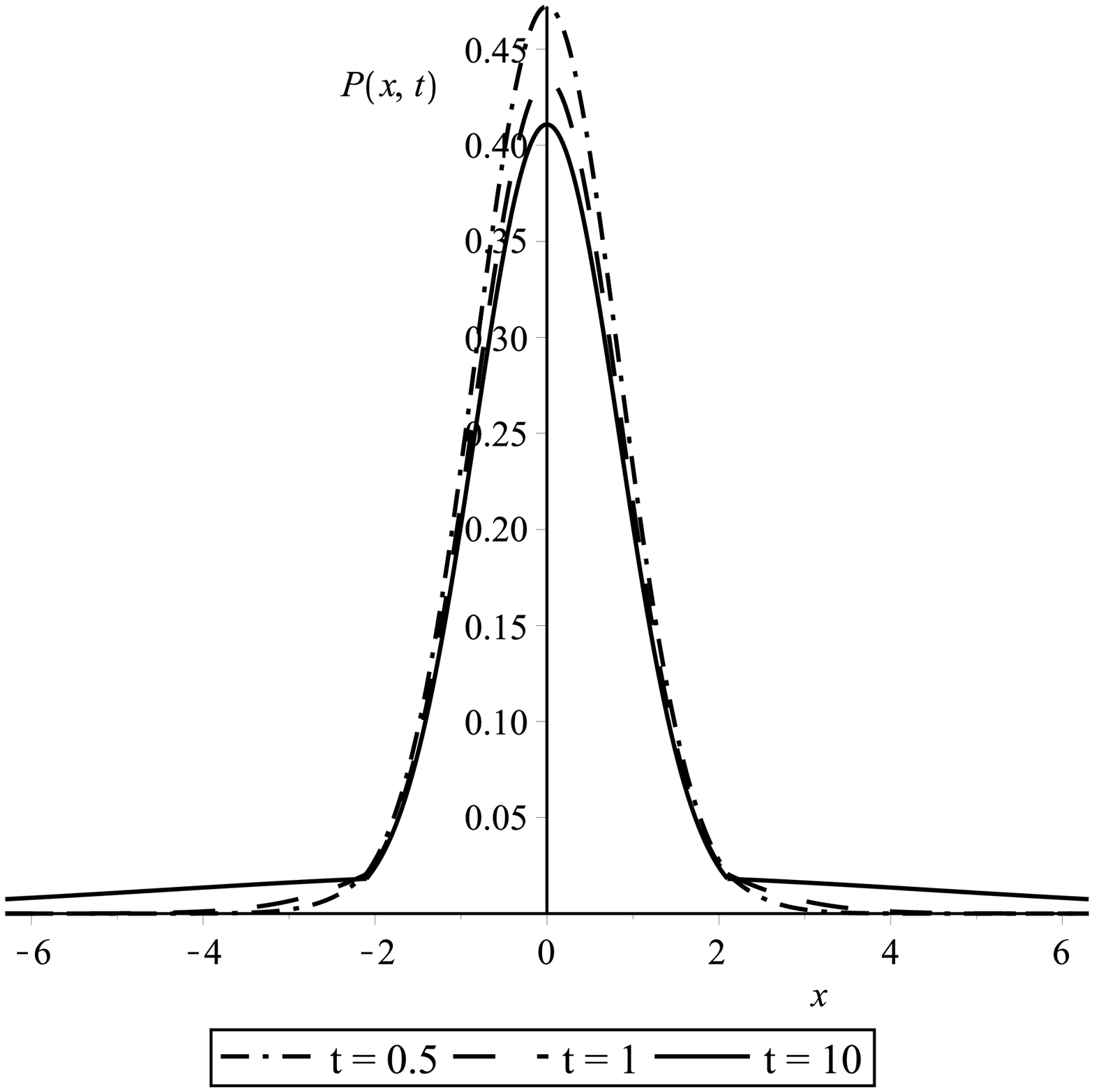}
}
\quad
{
\includegraphics[height=7cm]{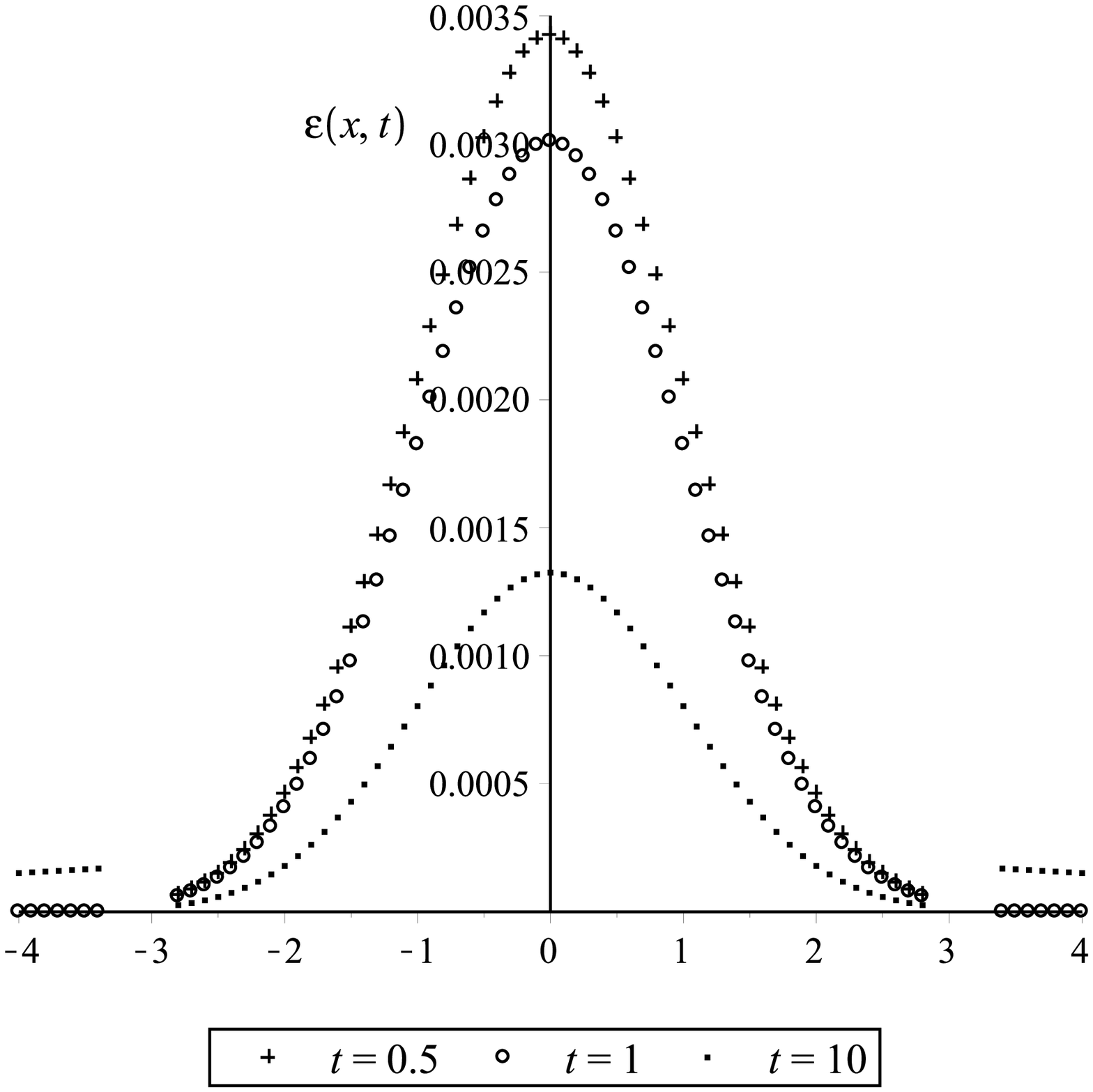}
}
\centerline{(a) \hspace{0.47\textwidth} (b)}
\caption{(a) ${P}({x,t})$, equation (\ref{eq18}) versus ${x}$ for different values of time for ${k} = 1.4$ and ${d} = 2.1$, ${Q} = 1$ and ${v_0} = 1.5$. (b) Estimated error $\varepsilon({x,t})$ for each time value.}
\label{fig5}
\end{figure}

Another example can be got by increasing the depth of the potential well for ${v_0} = {k^2 d^2 /4Q - k/2}  = 1.5$ with ${k} = 1.4$ and ${d} = 2.1$. The solution of (\ref{eq18}) under these conditions gives two terms of the series ($\lambda_0 = 0$ and $\lambda_1 = {k}$) and the probability distribution is represented in figure~\ref{fig5}, along with the associated error $\varepsilon({x,t})$ [equation (\ref{eq13})].

Comparison of figures~\ref{fig2}~(a) and \ref{fig5}~(a) shows that the curve of the probability distribution is smooth and the peak is more pronounced when the depth of the well is increased [figure~\ref{fig5}~(a). As previously discussed, the calculation of the error $\varepsilon ({x,t})$ shows a discontinuity at the points (${x} = \pm {d}$). It can be seen that, as time increases, there is a decrease in the calculated error indicating that the proposed solution is best for longer times.

\begin{figure}[!t]
\centerline{
\includegraphics[width=0.5\textwidth]{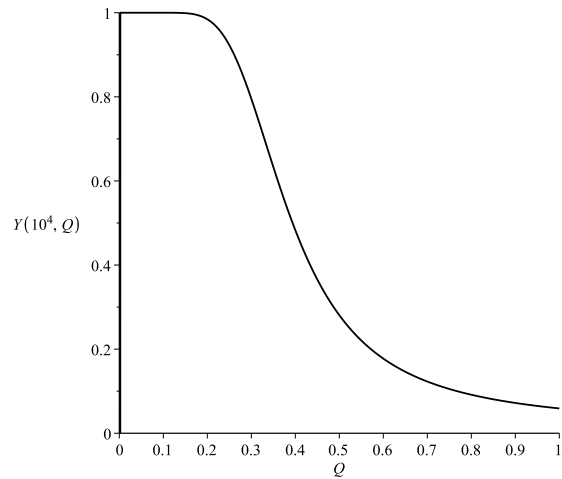}
}
\caption{Curve of ${Y}({t,Q})$, equation (\ref{eq12}) versus ${Q}$ for the truncated harmonic oscillator with two states within the well [equation (\ref{eq10}) with ${j} = 0$] and long time (${t} = 10^4$).}
\label{fig6}
\end{figure}

In figure~\ref{fig6}, the evolution of the number of particles as a function of the diffusion coefficient ${Q}$ for a long time, ${t} = 10^4$ is shown. Again, one can see a phase transition in the system. However, this transition is less sudden and its effects are noticeable at values of ${Q}$ higher than in the previous case (figure~\ref{fig4}). This effect is a result of the increased depth of the well, which makes the particle escape more difficult.


\section{A non-harmonic polynomial potential}
\label{sec4}

As a second example, the force associated with the system is assumed to be given by
\begin{equation}
\label{eq21}
f(x) = \left\{ \begin{array}{ll}
0, & \hbox{${ - d > x{\quad \text{and}\quad}x > d}$}, \\
- a{x^3} - b{x}, & \hbox{$- d \leqslant x \leqslant d$},
\end{array} \right.
\end{equation}
where ${a}$ and ${b}$ are two constants. On substituting this expression ${f}({x})$, equation (\ref{eq21}), in the Schr\"{o}dinger type equation, the effective potential, equation (\ref{eq4}), can be written as,
\begin{equation}
\label{eq22}
{V_\textrm{ef}}(x) = \left\{ \begin{array}{ll}
{v_0}, & \hbox{${ - d > x{\quad \text{and}\quad}x > d}$},\\
\frac{{{a^2}{x^6}}}{{4Q}} + \frac{{ab{x^4}}}{{2Q}} + \left( {\frac{{{b^2}}}{{4Q}} - \frac{{3a}}{2}} \right){x^2} - \frac{b}{2}, & \hbox{${ - d \leqslant x \leqslant d}$},
\end{array} \right.
\end{equation}
where the constants are chosen to ensure the continuity of the potential. For the region corresponding to ${x} > {d}$ and ${x} < -{d}$, the potential $V_\text{ef}({x})$ has a constant value ${v_0}$ equal to
\begin{equation}
\label{eq23}
{v_0} = \frac{{{a^2}{d^6}}}{{4Q}} + \frac{{ab{d^4}}}{{2Q}} + \left( {\frac{{{b^2}}}{{4Q}} - \frac{{3a}}{2}} \right){d^2} - \frac{b}{2}.
\end{equation}

\begin{figure}[!t]
\centerline{
\includegraphics[width=0.5\textwidth]{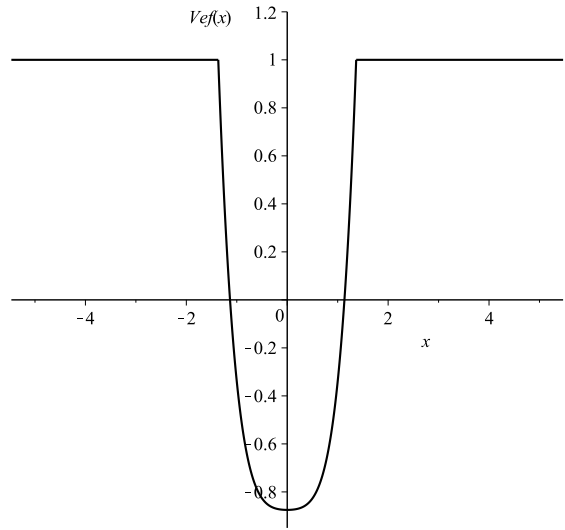}
}
\caption{Curve of potential (\ref{eq22}) versus ${x}$ for ${a}= 0.45$, ${b}= 1.75$, the depth ${v_0} = 1$, ${Q} =1$ and ${d} = 1.37$.} \label{fig7}
\end{figure}

Figure \ref{fig7} shows the curve of the partially confining potential given by equation (\ref{eq22}). The value of ${v_0}$ was fixed equal to 1 and the values of constants ${a}$ and ${b}$ were adjusted to allow just one minimum inside the potential well, the values used are ${a} = 0.45$ and ${b} = 1.75$. For these values of ${a}$ and ${b}$, if ${Q} = 1$, the intersection points are ${d} = \pm 1.37$.

In general, for non-harmonic polynomial potentials, the Schr\"{o}dinger equation (\ref{eq3}) has no exact/ana\-lytical solution. However, it is possible to determine part of the solution (partially soluble potential \cite{ref13,ref14}). In such cases, the approach introduced in this work can be used, approximating the solution for the original potential to that of the truncated potential and building ${P}({x,t})$ from function (\ref{eq10}), that is,
\begin{equation}
\label{eq24}
P(x,t) = \left\{ \begin{array}{ll}
{N_\textrm{I}}\frac{1}{{\sqrt {4Q\pi t} }}{\re^{ - {{{x^2}} \mathord{\left/
 {\vphantom {{{x^2}} {tQ}}} \right.
 \kern-\nulldelimiterspace} {tQ}}}}, & \hbox{$- d > x{\quad \text{and}\quad}x > d$},\\
{N_\textrm{I}}g(t)\sum\limits_{i = 0}^j {{a_i}{\phi _0}(x){\phi _i}(x)} {\re^{ - t\left| {{\lambda _i}} \right|}}, & \hbox{${ - d \leqslant x \leqslant d}$}.
\end{array} \right.
\end{equation}

In equation (\ref{eq24}) the functions $\phi_i({x})$ are chosen in order to satisfy the Schr\"{o}dinger equation (\ref{eq3}) with the potential (\ref{eq22}). The region of the well potential, described by equation (\ref{eq22}), does not give a general analytic/exact solution to all eigenfunctions $\phi_i({x})$. In this case, just the ground state is determined. However, depending on the well depth more eigenfunctions are necessary. Then, one possibility to get around this problem is to use other approximate methods, for example, the variational method \cite{ref15}.

Considering the potential characteristics studied and the parameters used (${a} = 0.45$, ${b} = 1.75$, ${Q} = 1$, ${d} = \pm 1.37$ and ${v_0} = 1$), there is only one state in the potential well region. Thus, only the first term of the series, should be considered in the region $-{d} \leqslant {x} \leqslant {d}$:
\begin{equation}
\label{eq25}
P(x,t) = {N_\textrm{I}}g(t){\phi _0}{(x)^2}{\re^{ - t{\lambda _0}}}.
\end{equation}

Then, in this example, when the potential is not truncated, just the ground state solution is analytically determined. In this case the eigenvalue ($\lambda_0$) is equal to 0 and the function $\phi_0^2({x})$ is the same adopted solution of stationary ground state of the Schr\"{o}dinger type equation when the potential well is infinite. Thus, the function to be used in equation (\ref{eq25}) is:
\begin{equation}
\label{eq26}
{\phi _0^2}{(x)} \propto {\exp\left\{ - \frac{a}{{4Q}}{x^4} - \frac{b}{{2Q}}{x^2}\right\}}.
\end{equation}

Therefore, the probability distribution for the force given by equation (\ref{eq21}) is represented as,
\begin{equation}
\label{eq27}
P(x,t) = \left\{ \begin{array}{ll}
{N_\textrm{I}}(t)\frac{1}{{\sqrt {4Q\pi t} }}{\re^{ - {{{x^2}} \mathord{\left/
 {\vphantom {{{x^2}} {tQ}}} \right.
 \kern-\nulldelimiterspace} {tQ}}}}, &  \hbox{$- d > x{\quad\text{and}\quad}x > d$},\\
{N_\textrm{I}}(t)g(t){\re^{ - \frac{a}{{4Q}}{x^4} - \frac{b}{{2Q}}{x^2}}},
& \hbox{ $- d \leqslant x \leqslant d$},
\end{array} \right.
\end{equation}
where the function ${g}({t})$  is given by
\begin{equation}
\label{eq28}
g(t) = \frac{{{\re^{ - {{{d^2}} \mathord{\left/
 {\vphantom {{{d^2}} {Qt}}} \right.
 \kern-\nulldelimiterspace} {Qt}}}}}}{{\sqrt {4\pi Qt} }}{\re^{{v_0}/Q}}
\end{equation}
and the normalization ${N_\textrm{I}}({t})$ is obtained from equation (\ref{eq11}),
\begin{equation}
\label{eq29}
{N_\textrm{I}}(t) = {{\left\{ {\frac{2}{{\sqrt {4\pi Qt} }}\int\limits_d^\infty  {{\re^{ - {{{x^2}} \mathord{\left/
 {\vphantom {{{x^2}} {tQ}}} \right.
 \kern-\nulldelimiterspace} {tQ}}}}} \rd x + g(t)\int\limits_{ - d}^d {{\re^{ - \frac{a}{{4Q}}{x^4} - \frac{b}{{2Q}}{x^2}}}} \rd x} \right\}^{-1}}}.
\end{equation}

\begin{figure}[!t]
\centering{
\includegraphics[height=7cm]{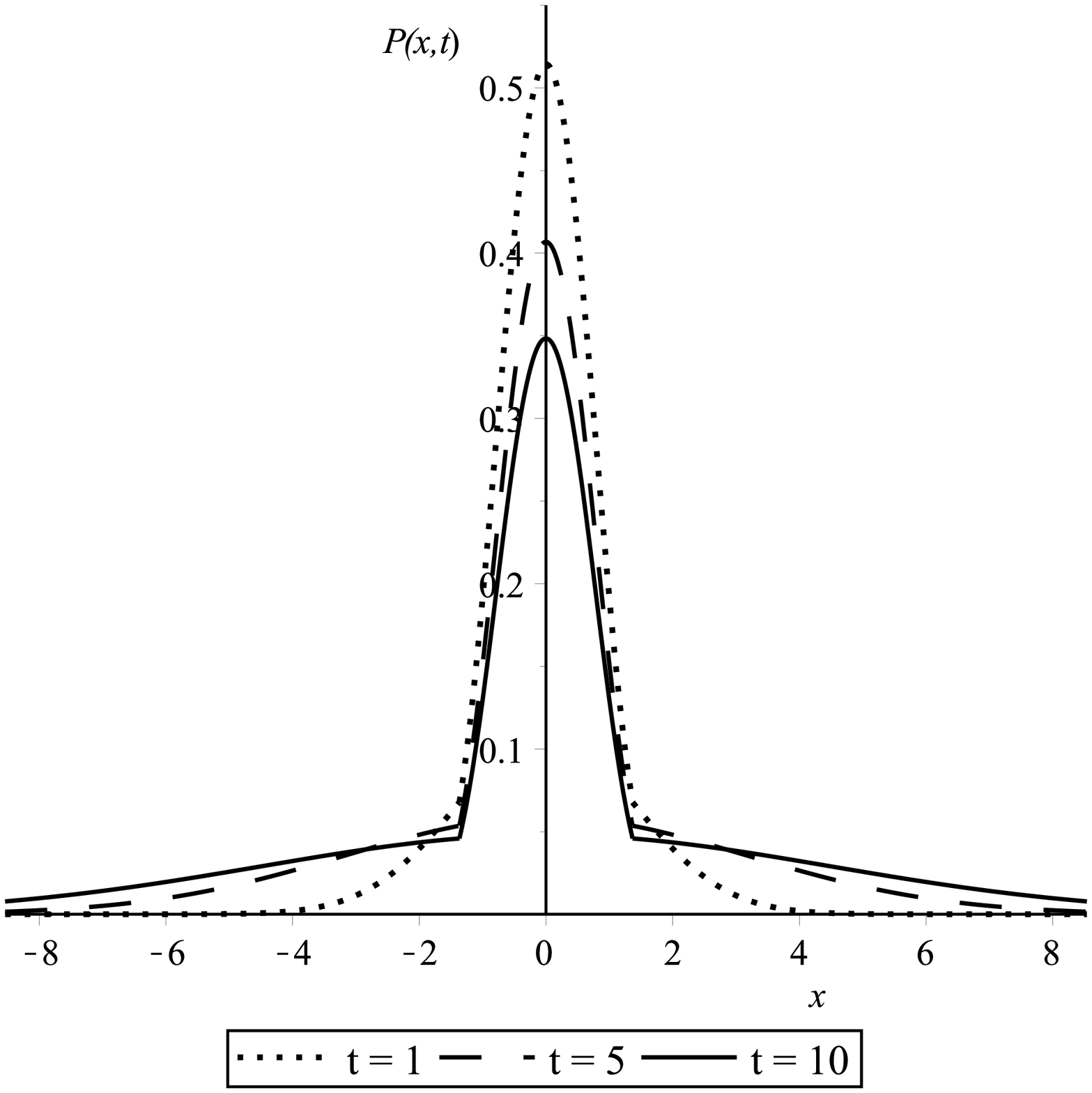}
}
\quad 
{
\includegraphics[height=7cm]{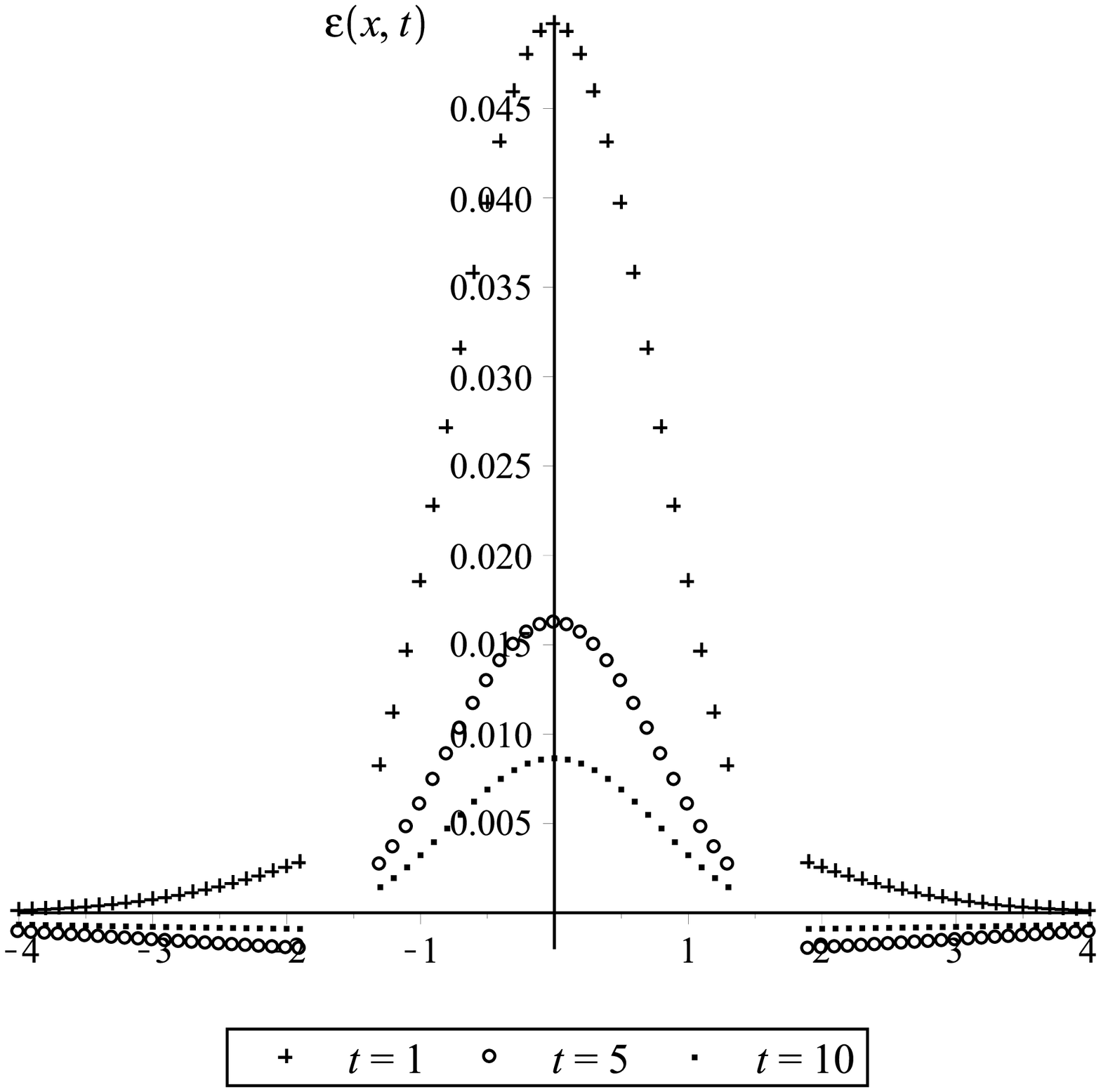}
}
\centerline{(a) \hspace{0.47\textwidth} (b)}
\caption{a) The probability distribution (\ref{eq27}) versus ${x}$, for truncated non-harmonic potential (\ref{eq22}). b) Error $\varepsilon ({x,t})$ of the approximate solution. The numeric constants used are ${a} = 0.45$, ${b} = 1.75$, ${Q} = 1$ and ${v_0} = 1$.}
\label{fig8}
\end{figure}

Figure~\ref{fig8} shows the probability distribution (\ref{eq27}) and the associated error $\varepsilon({x,t})$, equations (\ref{eq13}), for different values of time. The potential (\ref{eq22}) has the constants ${a}$ and ${b}$ equal to 0.45 and 1.75, respectively.

For numerical calculations, the depth of the well was fixed as ${v_0} = 1$, which implies a unique eigenvalue $\lambda_0 = 0$.  The interface points between the regions are
${d} = \pm 1.37$ and the value used for the diffusion coefficient to construct the curves shown in figure~\ref{fig8} is ${Q} = 1$.

It can be seen from figure~\ref{fig8}~(a) that the probability distribution ${P}({x,t})$
has a peak in the region of the potential well even for very long times. Initially, there is a very distinct peak probability (${t} = 1$) and, as time passes, this peak decreases and there is an increase in the width of the curve ${P}({x,t})$ at its base. The increased width of the probability distribution indicates the escape of the particles from the central region to the region of constant potential.

In the same way as for the harmonic case, the suggested solution is substituted in the Fokker-Planck equation to evaluate the error of the proposed method by determining the function $\varepsilon ({x,t})$ [figure~\ref{fig8}~(b)]. One can see that the approximate solution is better for large values of time than for shorter times. It is also noted that the largest error is in the region within the potential well and is smaller in the side regions where the potential is constant.

\section{Conclusion}
\label{sec5}

This paper presents an approximate analytical solution to the Fokker-Planck equation for partially confining potentials. The suggested solution corresponds to an adaptation of a previous proposal \cite{ref8} from the same authors. Here, there is suggested the removal of the Gaussian function from the region of potential well,  which permits a greater numerical accuracy of solution. This can be noted by using the expression $\varepsilon ({x,t})$, equation (\ref{eq13}).

In all the cases studied, following the initial condition ${P}({x}, 0) = \delta({x})$ and with the values of the constants as given in the examples, the probability distribution has a peak in the central region of the potential well. As the time increases, the curves ${P}({x,t})$ show a widening and a reduction in height. The calculation of $\varepsilon({x,t})$, equation (\ref{eq13}), as a way of assessing the accuracy of the approximate solutions in each example, indicates that the solutions have smaller errors  for longer values of time than for shorter times.

The approach outlined above allows for the study of a large number of problems whose solution proves difficult or impossible to obtain by other methods. For example, the truncated potentials discussed here could not be handled by the procedure given in reference \cite{ref9}, since it is not possible to get the exact/analytical solutions of the associated Schr\"{o}dinger type equation.

The suggested solution method has the advantage of being extended to classes of partially confining systems that do not have an exact analytical solution. Thus, an analytical expression, albeit approximate, of the probability distribution provides important information, allowing the study of a much larger number of systems. In addition to this, the use of the test function $\varepsilon({x,t})$ [equation~(\ref{eq13})] permits a quantitative measure of the accuracy of the result.

Analyzing the first example studied, i.e., the truncated harmonic potential with different depths, a transition phase can be identified involving the escape of particles from the well region of the potential. At low temperatures, the particles are trapped, while for higher temperatures the particles can escape. This escape leads to the emptying of the well. These results are very reliable, since they are obtained for long periods of time, a condition at which the proposed method turns out to be more accurate.

As a final remark, one observes that the approach introduced here can be addressed to the well known problem of the diffusion controlled escaping from a potential well \cite{ref16,ref17}. In this kind of problem, the calculation of the rate coefficients has a central importance \cite{ref18} and the calculation developed in the present work can be used to compute these quantities. Particularly, the escape rate problem is hard to analyze when the system is trapped in a potential well which correspond to the only point of minimum in the potential \cite{ref19}. In this context, the proposed function ${Y}({t,Q})$, equation (\ref{eq12}), can be useful.

\section*{Acknowledgements}
The authors acknowledge the financial support by the Brazilian agency CNPq (Proj. ESN \linebreak No.~233776/2014-1 and Proj. PDE No.~232865/2014-0) and the financial support of the Spanish MINECO (Project MTM2014-57129-C2-1-P) and Junta de Castilla y Le\'on (UIC 011).

\ukrainianpart

\title{Наближений розв'язок рівняння Фоккера-Планка}
\author{М.Т. Аройо\refaddr{label1,label2}, Е. Дріго Фільо\refaddr{label1,label3}}
\addresses{
\addr{label1} Інститут біології, UNESP --- державний університет штату Сан-Паулу, Сан-Жозе-ду-Ріу-Прету, Бразилія
\addr{label2} Інститут фізики, Університет м. Аугсбург, Німеччина
\addr{label3} Факультет теоретичної фізики, атомної  фізики та оптики та Інститут математики, \\ Вальядолідський університет, 47011 Вальядолід, Іспанія
}

\makeukrtitle

\begin{abstract}
У цій статті запропоновано наближений розв'язок  спеціального класу рівняння Фоккера-Планка. Розв'язок базується на зв'язку з рівнянням типу Шредингера з частково обмеженим і симетричним потенціалом. Щоб оцінити точність  розв'язку, запропоновано функцію похибок, яка отримана з оригінального рівняння Фоккера-Планка. Використовуючи запропонований метод, проаналізовано два приклади, а саме, утятий гармонічний потенціал і негармонічний поліном. Окрім цього, для утятого гармонічного потенціалу обговорено  поведінку системи в залежності від температури.
\keywords рівняння Фоккера-Планка, рівняння Шредингера, наближений розв'язок
\end{abstract}

\end{document}